\documentclass[runningheads]{llncs}
\usepackage{graphicx}
\usepackage{color}
\usepackage{url}
\usepackage{booktabs}
\usepackage{multirow}
\usepackage{subcaption}
\newcommand{\bhline}[1]{\noalign{\hrule height #1}}

\begin{document}
\title{Adaptive t-vMF Dice Loss \\ for Multi-class Medical Image Segmentation}
%
%
\author{Sota Kato \and Kazuhiro Hotta}
\authorrunning{S. Kato et al.}
%
\institute{Department of Electrical and Electronic Engineering, Meijo University, Japan\\ \email{150442030@ccalumni.meijo-u.ac.jp\\kazuhotta@meijo-u.ac.jp}}
%


\maketitle              
\begin{abstract}
Dice loss is widely used for medical image segmentation, and many improvement loss functions based on such loss have been proposed.
However, further Dice loss improvements are still possible.
In this study, we reconsidered the use of Dice loss and discovered that Dice loss can be rewritten in the loss function using the cosine similarity through a simple equation transformation.
Using this knowledge, we present a novel t-vMF Dice loss based on the t-vMF similarity instead of the cosine similarity.
Based on the t-vMF similarity, our proposed Dice loss is formulated in a more compact similarity loss function than the original Dice loss.
Furthermore, we present an effective algorithm that automatically determines the parameter $\kappa$ for the t-vMF similarity using a validation accuracy, called Adaptive t-vMf Dice loss. 
Using this algorithm, it is possible to apply more compact similarities for easy classes and wider similarities for difficult classes, and we are able to achieve an adaptive training based on the accuracy of the class.
Through experiments conducted on four datasets using a five-fold cross validation, we confirmed that the Dice score coefficient (DSC) was further improved in comparison with the original Dice loss and other loss functions.
Our code is available at \url{https://github.com/usagisukisuki/Adaptive_t-vMF_Dice_loss}.

\end{abstract}
%
%

\section{Introduction}
In recent years, segmentation tasks that assign the class label to each pixel in an image have become important in the field of medicine\cite{b1,b2,b3} and biology\cite{b4,b5}.
Although various network architectures have been proposed to improve the segmentation accuracy\cite{b6,b7,b8,b9}, during the loss function term, Dice loss \cite{b10} has been used in most cases.
Dice loss brings a dataset consisting of all positive examples predicted by a model closer to a dataset consisting of the ground truth.
The Dice score coefficient (DSC) is a measure of overlap that is widely used to assess the segmentation performance when a ground truth is available, and the Dice loss can be optimized the DSC directly.
Then it is a powerful way to achieve semantic segmentation.
It is also effective for datasets with many negative samples in an image because it calculates the loss as a percentage of the region that can be detected.
In recent studies, various loss functions based on Dice loss have been proposed to further improve the DSC\cite{b11,b12,b13,b14,b15,b16}.
However, in the case of a multi-class segmentation task, the areas of predicted regions are often different for each class, and tend to have an imbalanced segmentation.
As a result, the DSC of classes with small regions tends to be lower than that of classes with large regions.
Conventional loss functions based on Dice loss have been unable to solve this problem.

Therefore, we reconsider a Dice loss equation and normalize both the dataset consisting of the predicted regions and the dataset consisting of ground truth regions to eliminate the difference in regions between classes.
As a result, the original Dice loss is able to rewrite in the loss function using the cosine similarity based on the dot product of the features.
In recent years, the t-vMF similarity \cite{b20}, which is an extended method for the cosine similarity, has also been proposed. 
In this study, we present t-vMF Dice loss based on the t-vMF similarity.
The t-vMF Dice loss is formulated in a more compact similarity loss function than the original Dice loss.
Furthermore, we present an algorithm that automatically determines the parameter $\kappa$ for the t-vMF similarity using the Dice coefficient for the validation set, which we named Adaptive t-vMF Dice loss. 
Using Adaptive t-vMF Dice loss, it is possible to use more compact similarities for easy classes, and wider similarities for difficult classes.

We evaluated our loss function experimentally on four datasets.
From the experiment results, although we changed only the loss function, we confirmed that the proposed method was significantly improved in comparison with the Dice loss and conventional loss functions for the image segmentation datasets.

This paper is organized as follows. 
Section 2 describes the related works of loss functions for segmentation. 
Section 3 explains the details of the proposed method. 
Section 4 presents the experimental results. 
Finally, we describe our summary and future works in Section 5.

The main contributions of this paper are as follows:
 \begin{itemize}
  \item[$\bullet$] Through a simple formula transformation, we discovered that the original Dice loss was able to rewrite in the loss function using the cosine similarity.
  \item[$\bullet$] We proposed two types of novel loss functions, called t-vMF Dice loss and Adaptive t-vMF Dice loss. The t-vMF Dice loss is formulated in a more compact similarity loss function than the conventional Dice loss, and Adaptive t-vMF Dice loss is able to use more compact similarities for easy classes and wider similarities for difficult classes. 
 \end{itemize}

\section{Related work}
In recent studies, Dice loss has been the most frequently used loss function in medical image segmentation.
Dice loss is a loss function that brings a dataset consisting of all positive instances predicted by a model closer to a dataset consisting of the ground truth, and is a powerful method for achieving a semantic segmentation because it can directly optimize the Dice score coefficient (DSC).
Various loss functions based on the Dice loss have been proposed \cite{b11,b12,b13,b14,b15,b16}. 
Li et al.\cite{b11} proposed a generalized Dice loss that
is able to use the class re-balancing properties of the generalized Dice overlap and achieve a robust and accurate deep learning loss function for an unbalanced segmentation.
Abraham et al.\cite{b16} used the Tversky similarity index, which is a generalization of the Dice score that can flexibly balance between false positives and false negatives, and then proposed the focal Tversky loss.

Although various loss functions for a semantic segmentation have been proposed, the imbalanced datasets for multi-class segmentation have yet to be sufficiently solved.
Our proposed loss functions have a more compact similarity, and we can change the compactness of the similarity for each class. 
A high performance is possible in comparison with conventional loss functions even if we work on imbalanced segmentation tasks.

\section{Methodology}
This section describes the analysis of the original Dice loss and our proposed loss function.
In Section 3.1, we analyze the equation of the original Dice loss, and present the potential for improving the level of accuracy.
In Section 3.2, we present two novel loss functions called t-vMF Dice loss and Adaptive t-vMF Dice loss.

\subsection{Analysis of the original Dice loss}
Equation (1) shows the original Dice loss.
\begin{eqnarray}
  Dice\ loss &=& \frac{1}{C}\sum_{i=1}^C(1 - \frac{2\sum_{n}A_{in} B_{in}+\gamma}{\sum_{n} A_{in}^2 + \sum_{n} B_{in}^2 +\gamma})
\end{eqnarray}
where $C$ indicates the number of classes, $n$ indicates the number of class samples, $A_{in}$ indicates those vectors containing all positive examples predicted by a specific model, and $B_{in}$ indicates the vectors containing all positive examples of the ground truth in the dataset.
For the purposes of smoothing, it is common to add a $\gamma$ factor to both the nominator and denominator, and in general, $\gamma = 1$.
Then, when the cosine similarity is defined, equation (1) is expanded into equation (2).

\begin{eqnarray}
      Dice\ loss &=& \frac{1}{C}\sum_{i=1}^C(1 - \frac{\sum_{n}A_{in} B_{in}}{\sqrt{\sum_{n}A_{in}^2}\sqrt{\sum_{n}B_{in}^2}}\cdot\frac{2\sqrt{\sum_{n}A_{in}^2}\sqrt{\sum_{n}B_{in}^2}}{\sum_{n} A_{in}^2 + \sum_{n} B_{in}^2})\nonumber \\
      &=& \frac{1}{C}\sum_{i=1}^C(1 -\cos \theta_i\cdot\frac{2\sqrt{\sum_{n}A_{in}^2}\sqrt{\sum_{n}B_{in}^2}}{\sum_{n} A_{in}^2 + \sum_{n} B_{in}^2})
\end{eqnarray}
where we have $\gamma=0$ to simplify the equation.
In equation (2), we confirm that the original Dice loss contains the cosine similarity.
However, the lengths of two vectors $A_{in}$ and $B_{in}$ vary depending on each class, which causes an imbalanced segmentation.
Then, we normalize two vectors of $A_{in}$ and $B_{in}$ for each class to eliminate the difference in regions between classes.
When we define $\sqrt{\sum_{n}A_{in}^2}=\sqrt{\sum_{n}B_{in}^2}=1$, equation (2) can be rewritten into a new equation using the cosine similarity in equation (3).

\begin{eqnarray}
      Dice\ loss_{norm} 
      &=& \frac{1}{C}\sum_{i=1}^C(1 - \cos \theta_{i})
\end{eqnarray}
where $C$ is the number of classes, and $\cos \theta_{i}=\sum_{n}A_{in} B_{in}$.
This is compatible with $\cos \theta_{i} \in [0, +1]$ because the output used by the Dice loss is normalized using the sigmoid or softmax function.
The original Dice loss is then a loss function that maintains the cosine similarity between the set of values predicted by the model and the set of ground truth values at close to 1.

\subsection{Adaptive t-vMF Dice loss}
\begin{figure}[t]
    \centering
    \begin{tabular}{cc}
      \begin{minipage}{0.45\hsize}
        \includegraphics[scale=0.44]{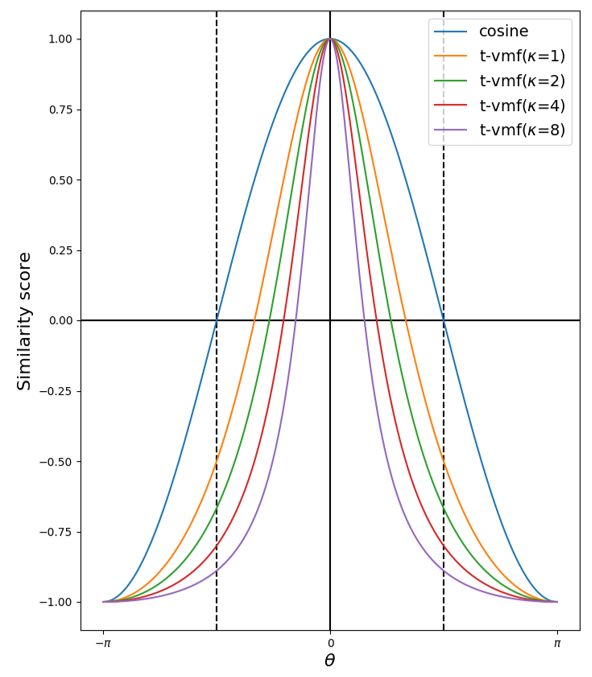}
        \subcaption{}
    \end{minipage}%
    \begin{minipage}{0.45\hsize}
        \includegraphics[scale=0.50]{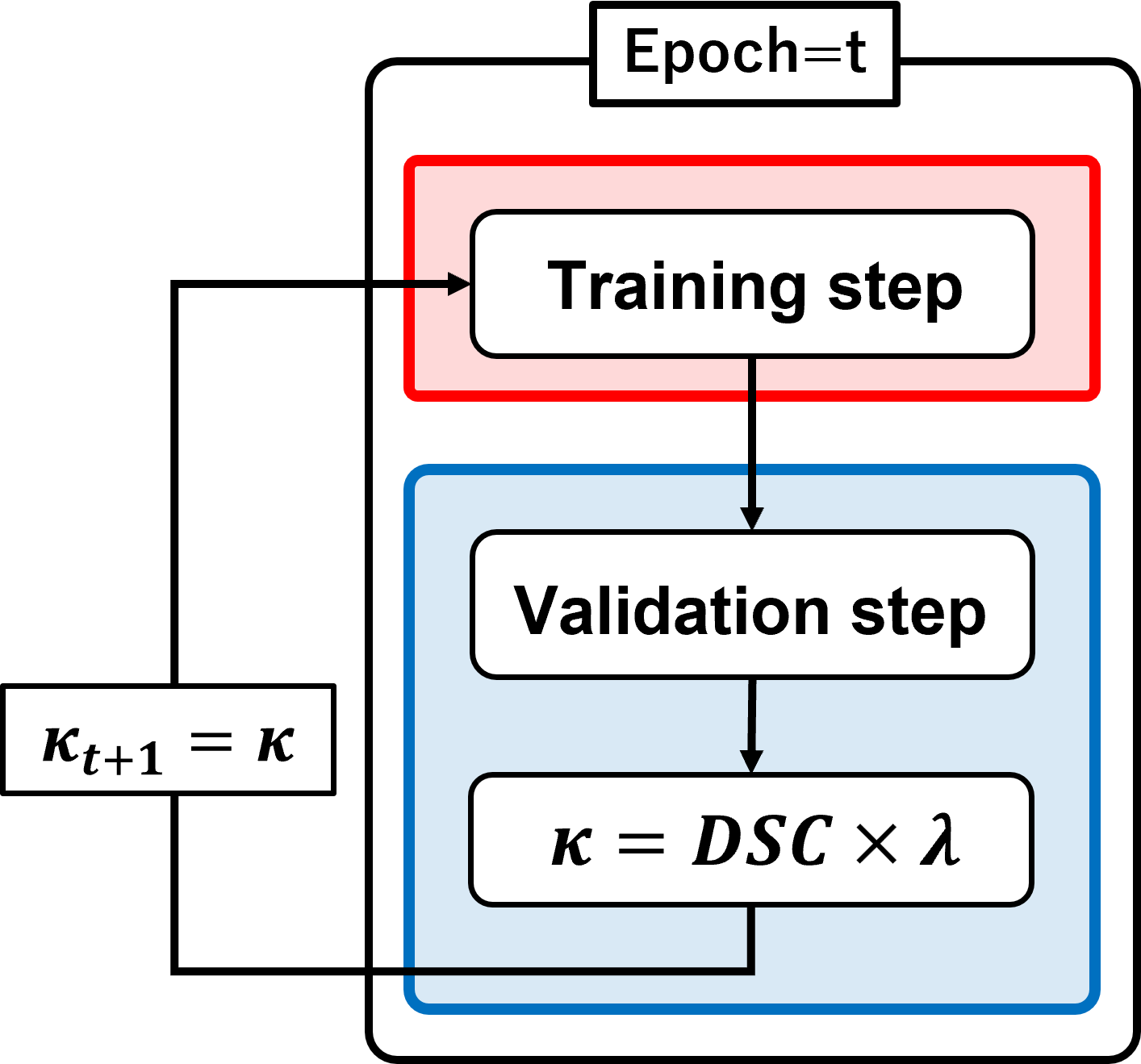}
        \subcaption{}
    \end{minipage}%
    \end{tabular}
    \caption{ (a) Comparison of cosine similarity and t-vMF-based similarities. (b) Overview of the training and validation flows using Adaptive t-vMF Dice loss.}
\end{figure}

In Section 3.1, we confirmed that the original Dice loss is able to rewrite to the equation using the cosine similarity.
Therefore, we adopted a novel approach applying the t-vMF similarity to the Dice loss.
The t-vMF similarity, which is an extension of the cosine similarity was proposed in \cite{b20}.
Equation (4) shows the t-vMF similarity, and Fig. 1 (a) shows a comparison of the cosine similarity and the t-vMF based similarities.
\begin{eqnarray}
  \phi_{t}(\cos \theta; \kappa) &=& \frac{1+\cos \theta}{1+\kappa(1-\cos \theta)}-1
\end{eqnarray}
By changing the parameter $\kappa$, it is possible to freely change the range of similarities and achieve a more compact similarity than the cosine similarity using a large $\kappa$. 
Here, $\kappa$ = 0 is exactly the same as the original cosine similarity, $\phi_{t}(\cos \theta; 0) =  \cos \theta$ without a treatment about practical computation.
With this compact similarity and equation (3), we present a novel t-vMF Dice loss in equation (5).
\begin{eqnarray}
  t\mathchar`-vMF\ Dice\ loss &=& \frac{1}{C}\sum_{i=1}^C(1 - \phi_{t}(\cos \theta_{i}; \kappa))^2
\end{eqnarray}
when $\cos \theta \in [0, +1]$ and $\phi_{t}(\cos \theta; \kappa=1) \in [-0.5, +1]$, because the similarity takes a negative value, we use the mean squared error loss (MSE loss) to place the similarity closer to 1 according to the original Dice loss.
In equation (5), the novel Dice loss has a more compact similarity than the conventional Dice loss.

However, the parameter $\kappa$ is a fixed value in all classes, and we should use a suitable $\kappa$ for each class.
Therefore, we add an effective algorithm to the t-vMF Dice loss to adaptively determine $\kappa$, and present Adaptive t-vMF Dice loss.
As a measure for determining the parameter $\kappa$, we use the DSC for the validation set. 
Fig. 1 (b) shows the flow of training and validation in the Adaptive t-vMF Dice loss.
After we evaluate the DSC of each class during validation step, we multiply the class DSC by $\lambda$ to obtain the $\kappa$ in the next epoch.
Here, $\lambda$ is a hyper parameter that determines the upper limit of $\kappa$, which we chose based on \cite{b20}.
The initial value of $\kappa$ at the first epoch is zero.
Using Adaptive t-vMF Dice loss, it possible to dynamically choose the most suitable $\kappa$ for each class and achieve more compact similarities for an easy class and wider similarities for difficult classes.

\section{Experiments}
\begin{table}[t]
    \centering
    \caption{Comparison results for binary segmentations.}
    \scalebox{0.9}{
    \begin{tabular*}{12cm}{@{\extracolsep{\fill}}rcccc} \bhline{1.0pt}
    \multicolumn{1}{r}{} & \multicolumn{2}{c}{CVC-ClinicDB} & \multicolumn{2}{c}{Kvasir-SEG}\\
    \cmidrule(lr){2-3}%
    \cmidrule(lr){4-5}%
    \multicolumn{1}{r}{Loss function} & \multicolumn{1}{c}{U-Net} & \multicolumn{1}{c}{TransUNet} & \multicolumn{1}{c}{U-Net} & \multicolumn{1}{c}{TrasUNet}\\
    \hline
    Dice loss &82.02\tiny{±4.94}&91.41\tiny{±3.93}&90.34\tiny{±0.71}&93.50\tiny{±0.49}\\
        Generalised Dice loss &83.37\tiny{±2.53}&91.01\tiny{±4.37}&88.70\tiny{±0.51}&93.21\tiny{±0.51}\\
        Noise-robust Dice loss &83.14\tiny{±4.71}&\textcolor{red}{\textbf{91.91\tiny{±3.57}}}&90.34\tiny{±0.58}&93.53\tiny{±0.49}\\
        WS Dice loss &82.41\tiny{±3.10}&90.49\tiny{±4.82}&90.30\tiny{±0.54}&92.67\tiny{±0.79}\\
        Focal Dice loss &84.38\tiny{±4.34}&91.13\tiny{±3.94}&90.98\tiny{±0.51}&93.50\tiny{±0.54}\\
        $\beta FL-log (Dice)$ loss  &86.57\tiny{±3.39}&89.22\tiny{±5.48}&91.80\tiny{±0.54}&92.22\tiny{±0.57}\\
        Focal Tversky loss &83.39\tiny{±4.12}&91.81\tiny{±3.48}&90.61\tiny{±0.20}&93.42\tiny{±0.37}\\
        \hline
        t-vMF Dice loss ($\kappa$=2)&83.05\tiny{±3.44}&90.51\tiny{±4.49}&91.01\tiny{±0.42}&93.55\tiny{±0.30}\\
        t-vMF Dice loss ($\kappa$=32)&86.77\tiny{±5.10}&90.46\tiny{±3.76}&92.31\tiny{±0.30}&93.39\tiny{±0.36}\\
        t-vMF Dice loss ($\kappa$=128)&88.16\tiny{±2.43}&90.76\tiny{±4.39}&\textcolor{red}{\textbf{92.45\tiny{±0.10}}}&93.39\tiny{±0.36}\\
        Adaptive t-vMF Dice loss ($\lambda$=2) &82.73\tiny{±3.59}&91.32\tiny{±3.76}&90.66\tiny{±0.57}&\textcolor{red}{\textbf{93.55\tiny{±0.30}}}\\
        Adaptive t-vMF Dice loss ($\lambda$=32) &87.39\tiny{±4.56}&90.71\tiny{±4.93}&92.18\tiny{±0.30}&92.66\tiny{±0.57}\\
        Adaptive t-vMF Dice loss ($\lambda$=128) &\textcolor{red}{\textbf{88.68\tiny{±3.87}}}&90.34\tiny{±4.21}&92.24\tiny{±0.32}&92.60\tiny{±0.32}\\
    \bhline{1.0pt} 
    \end{tabular*}
    }
\end{table}
\begin{table}[t]
    \centering
    \caption{Comparison results for ACDC dataset.}
    \scalebox{0.61}{
    \begin{tabular*}{19cm}{@{\extracolsep{\fill}}rcccccccc} \bhline{1.0pt}
    \multicolumn{1}{r}{} & \multicolumn{4}{c}{U-Net} & \multicolumn{4}{c}{TransU-Net}\\
    \cmidrule(lr){2-5}%
    \cmidrule(lr){6-9}%
    \multicolumn{1}{r}{Loss function} & Average DSC & RV & MYO & LV & Average DSC & RV & MYO & LV\\
    \hline
        Dice loss &92.21\tiny{±1.30}&89.92\tiny{±2.72}&84.37\tiny{±1.94}&94.76\tiny{±1.24}&92.72\tiny{±0.80}&91.72\tiny{±1.46}&84.58\tiny{±1.33}&94.80\tiny{±1.49}\\
        Generalised Dice loss  &92.42\tiny{±1.46}&90.37\tiny{±3.05}&84.70\tiny{±1.70}&94.84\tiny{±1.54}&93.01\tiny{±9.65}&92.15\tiny{±1.35}&85.21\tiny{±1.14}&94.89\tiny{±1.15}\\
        Noise-robust Dice loss  &92.40\tiny{±1.30}&90.29\tiny{±3.01}&84.64\tiny{±1.72}&94.88\tiny{±1.35}&92.84\tiny{±0.73}&91.95\tiny{±1.57}&84.67\tiny{±1.24}&94.96\tiny{±1.24}\\
        WS Dice loss &90.57\tiny{±1.24}&87.08\tiny{±3.08}&81.03\tiny{±2.01}&94.51\tiny{±1.41}&91.03\tiny{±0.95}&88.72\tiny{±2.57}&81.16\tiny{±2.44}&94.55\tiny{±1.16}\\
        Focal Dice loss &92.82±\tiny{1.29}&90.56\tiny{±2.91}&85.68\tiny{±1.38}&95.27\tiny{±1.18}&92.46\tiny{±0.70}&91.49\tiny{±1.47}&83.91\tiny{±1.20}&94.64\tiny{±1.28}\\
        $\beta FL-log (Dice)$ loss  &92.53\tiny{±1.37}&90.47\tiny{±2.81}&84.94\tiny{±1.55}&94.93\tiny{±1.48}&92.94\tiny{±0.63}&91.96\tiny{±1.67}&84.91\tiny{±1.11}&95.08\tiny{±0.88}\\
        Focal Tversky loss &92.44\tiny{±1.47}&90.14\tiny{±3.50}&84.57\tiny{±1.81}&95.31\tiny{±1.03}&92.74\tiny{±0.47}&91.71\tiny{±0.87}&84.48\tiny{±0.75}&95.02\tiny{±0.93}\\
        \hline
        t-vMF Dice loss ($\kappa$=2) &92.75\tiny{±1.37}&90.38\tiny{±2.94}&85.78\tiny{±1.45}&95.07\tiny{±1.40}&92.72\tiny{±0.78}&91.63\tiny{±1.42}&84.55\tiny{±1.19}&94.92\tiny{±1.38}\\
        t-vMF Dice loss ($\kappa$=32) &88.99\tiny{±1.22}&86.57\tiny{±3.10}&77.47\tiny{±1.53}&92.45\tiny{±1.30}&92.21\tiny{±1.30}&89.92\tiny{±2.72}&84.37\tiny{±1.94}&94.76\tiny{±1.24}\\
        t-vMF Dice loss ($\kappa$=128) &93.62\tiny{±0.71}&92.35\tiny{±0.96}&86.62\tiny{±1.08}&95.69\tiny{±1.09}&55.93\tiny{±10.07}&48.48\tiny{±22.24}&75.47\tiny{±18.69}&0.00\tiny{±0.00}\\
        Adaptive t-vMF Dice loss ($\lambda$=2) &92.79\tiny{±1.36}&90.62\tiny{±2.82}&85.60\tiny{±1.73}&95.15\tiny{±1.39}&92.79\tiny{±0.55}&91.69\tiny{±1.32}&84.70\tiny{±0.74}&94.98\tiny{±1.08}\\
        Adaptive t-vMF Dice loss ($\lambda$=32) &\textcolor{red}{\textbf{93.68\tiny{±0.93}}}&\textcolor{red}{\textbf{92.61\tiny{±1.46}}}&\textcolor{red}{\textbf{86.73\tiny{±1.40}}}&95.56\tiny{±1.20}&93.29\tiny{±0.78}&92.22\tiny{±1.39}&85.66\tiny{±1.11}&95.47\tiny{±1.04}\\
        Adaptive t-vMF Dice loss ($\lambda$=128) &93.50\tiny{±1.08}&92.07\tiny{±1.73}&86.45\tiny{±1.53}&\textcolor{red}{\textbf{95.70\tiny{±1.17}}}&\textcolor{red}{\textbf{93.43\tiny{±0.66}}}&\textcolor{red}{\textbf{92.50\tiny{±1.00}}}&\textcolor{red}{\textbf{85.80\tiny{±1.20}}}&\textcolor{red}{\textbf{95.62\tiny{±1.11}}}\\
    \bhline{1.0pt} 
    \end{tabular*}
    }
\end{table}
\begin{table}[t]
    \centering
    \caption{Comparison results for Synapse multi-organ segmentation dataset.}
    \scalebox{0.54}{
    \begin{tabular}{crccccccccc} \bhline{1.0pt}
    Networks & Loss function & Average DSC & Aorta & Gallbladder & Kidney(L) & Kidney(R) & Liver & Pancreas & Spleen & Stomach \\
    \hline
    \multirow{11}{*}{U-Net}&Dice loss  &70.57\tiny{±7.38}&78.08\tiny{±4.88}&58.72\tiny{±12.16}&65.42\tiny{±6.96}&59.73\tiny{±11.91}&87.03\tiny{±5.52}&48.89\tiny{±8.83}&74.44\tiny{±21.86}&63.37\tiny{±6.63}\\
    &Generalised Dice loss  &66.53\tiny{±8.80}&77.77\tiny{±7.66}&27.30\tiny{±33.45}&64.97\tiny{±9.07}&60.89\tiny{±12.81}&87.59\tiny{±4.59}&45.44\tiny{±5.75}&73.70\tiny{±21.21}&61.67\tiny{±5.36}\\
    &Noise-robust Dice loss &64.74\tiny{±6.45}&77.96\tiny{±6.70}&9.81\tiny{±19.62}&65.82\tiny{±6.92}&61.80\tiny{±10.62}&87.42\tiny{±4.71}&46.18\tiny{±7.57}&71.92\tiny{±21.30}&62.32\tiny{±3.66}\\
    &Focal Dice loss &68.97\tiny{±6.86}&74.72\tiny{±6.38}&59.67\tiny{±8.43}&65.36\tiny{±8.73}&59.73\tiny{±12.46}&84.01\tiny{±5.47}&46.04\tiny{±7.20}&71.28\tiny{±20.15}&60.62\tiny{±8.73}\\
    &$\beta FL-log (Dice)$ loss  &67.63\tiny{±7.97}&77.16\tiny{±5.35}&34.02\tiny{±29.71}&65.62\tiny{±13.70}&59.72\tiny{±12.94}&86.64\tiny{±6.45}&50.01\tiny{±6.51}&74.16\tiny{±21.55}&61.93\tiny{±4.63}\\
    &Focal Tversky loss &64.41\tiny{±5.80}&75.94\tiny{±5.98}&12.53\tiny{±25.07}&64.29\tiny{±11.03}&62.52\tiny{±11.43}&84.52\tiny{±5.04}&45.00\tiny{±3.35}&72.48\tiny{±21.40}&63.14\tiny{±5.47}\\
    \cmidrule(r){2-11}%
    &t-vMF Dice loss ($\kappa$=32)&49.76\tiny{±3.66}&16.80\tiny{±25.22}&0.00\tiny{±0.00}&69.76\tiny{±7.38}&65.76\tiny{±8.42}&88.12\tiny{±4.06}&6.14\tiny{±4.83}&75.43\tiny{±20.00}&59.96\tiny{±3.47}\\
    &t-vMF Dice loss ($\kappa$=256)&17.75\tiny{±1.00}&0.25\tiny{±0.11}&0.02\tiny{±0.02}&0.71\tiny{±0.85}&0.15\tiny{±0.13}&82.45\tiny{±5.05}&0.01\tiny{±0.01}&0.05\tiny{±0.03}&22.70\tiny{±6.22}\\
    &Adaptive t-vMF Dice loss ($\lambda$=32)&73.81\tiny{±6.13}&\textcolor{red}{\textbf{81.67\tiny{±4.68}}}&\textcolor{red}{\textbf{62.33\tiny{±14.70}}}&71.56\tiny{±5.39}&\textcolor{red}{\textbf{68.44\tiny{±8.74}}}&\textcolor{red}{\textbf{88.70\tiny{±4.99}}}&\textcolor{red}{\textbf{50.39\tiny{±8.52}}}&75.53\tiny{±20.06}&66.21\tiny{±5.09}\\
    &Adaptive t-vMF Dice loss ($\lambda$=256)&\textcolor{red}{\textbf{74.22\tiny{±7.93}}}&81.22\tiny{±4.95}&61.65\tiny{±16.93}&\textcolor{red}{\textbf{74.67\tiny{±5.40}}}&68.33\tiny{±12.78}&89.04\tiny{±5.60}&49.99\tiny{±7.21}&\textcolor{red}{\textbf{76.55\tiny{±24.62}}}&\textcolor{red}{\textbf{67.03\tiny{±5.60}}}\\
    \hline
    \multirow{11}{*}{TransUNet}&Dice loss &76.46\tiny{±6.72}&83.19\tiny{±5.06}&55.58\tiny{±10.11}&77.97\tiny{±9.08}&77.24\tiny{±8.75}&90.55\tiny{±4.90}&53.20\tiny{±7.23}&78.67\tiny{±19.99}&72.17\tiny{±1.86}\\
    &Generalised Dice loss &63.42\tiny{±3.36}&14.94\tiny{±29.88}&56.12\tiny{±8.11}&79.41\tiny{±10.19}&78.42\tiny{±10.30}&90.36\tiny{±4.56}&0.00\tiny{±0.00}&78.90\tiny{±19.52}&73.13\tiny{±2.72}\\
    &Noise-robust Dice loss  &60.42\tiny{±5.32}&0.00\tiny{±0.00}&52.06\tiny{±8.08}&76.16\tiny{±10.44}&74.88\tiny{±11.59}&90.43\tiny{±4.78}&0.00\tiny{±0.00}&79.12\tiny{±20.00}&71.67\tiny{±2.97}\\
    &Focal Dice loss &76.01\tiny{±5.89}&82.78\tiny{±4.06}&55.07\tiny{±5.32}&78.48\tiny{±8.14}&76.60\tiny{±9.39}&89.82\tiny{±4.95}&52.34\tiny{±6.71}&78.27\tiny{±20.09}&71.18\tiny{±3.15}\\
    &$\beta FL-log (Dice)$ loss  &76.23\tiny{±6.22}&83.73\tiny{±3.30}&45.87\tiny{±25.17}&80.21\tiny{±9.14}&79.20\tiny{±9.29}&91.10\tiny{±4.72}&52.52\tiny{±6.46}&79.97\tiny{±19.15}&73.87\tiny{±4.58}\\
    &Focal Tversky loss &60.81\tiny{±4.83}&67.38\tiny{±2.87}&10.01\tiny{±20.03}&64.93\tiny{±7.10}&73.73\tiny{±8.79}&80.04\tiny{±4.82}&37.35\tiny{±5.58}&65.16\tiny{±13.46}&59.42\tiny{±4.76}\\
    \cmidrule(r){2-11}%
    &t-vMF Dice loss ($\kappa$=32)&63.73\tiny{±3.53}&18.08\tiny{±32.93}&32.58\tiny{±26.64}&62.80\tiny{±31.59}&78.71\tiny{±9.58}&91.39\tiny{±4.83}&49.66\tiny{±5.53}&82.35\tiny{±18.15}&73.21\tiny{±2.97}\\
    &t-vMF Dice loss ($\kappa$=256)&24.22\tiny{±1.87}&0.25\tiny{±0.40}&4.45\tiny{±8.89}&7.53\tiny{±13.04}&31.63\tiny{±20.02}&90.96\tiny{±4.68}&7.32\tiny{±6.70}&0.05\tiny{±0.07}&\textcolor{red}{\textbf{75.75\tiny{±5.81}}}\\
    &Adaptive t-vMF Dice loss ($\lambda$=32)&\textcolor{red}{\textbf{78.27\tiny{±5.54}}}&\textcolor{red}{\textbf{85.64\tiny{±3.34}}}&\textcolor{red}{\textbf{62.10\tiny{±7.97}}}&79.47\tiny{±10.14}&78.54\tiny{±10.85}&91.47\tiny{±4.39}&52.25\tiny{±8.82}&\textcolor{red}{\textbf{81.68\tiny{±18.39}}}&73.65\tiny{±3.64}\\
    &Adaptive t-vMF Dice loss ($\lambda$=256)&77.10\tiny{±5.06}&84.46\tiny{±3.77}&48.47\tiny{±25.79}&\textcolor{red}{\textbf{80.15\tiny{±8.38}}}&\textcolor{red}{\textbf{78.90\tiny{±8.40}}}&\textcolor{red}{\textbf{92.06\tiny{±4.74}}}&\textcolor{red}{\textbf{53.79\tiny{±8.08}}}&80.97\tiny{±18.93}&75.45\tiny{±1.34}\\
    \bhline{1.0pt} 
    \end{tabular}
    }
\end{table}
\begin{figure*}[t]
    \centering
    \begin{tabular}{cccc}
      \begin{minipage}{0.23\hsize}
        \centering
        \includegraphics[scale=0.34, angle=180]{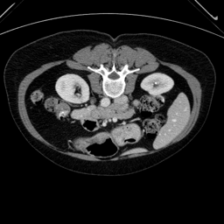}
    \end{minipage}%
    \begin{minipage}{0.23\hsize}
        \centering
        \includegraphics[scale=0.34, angle=180]{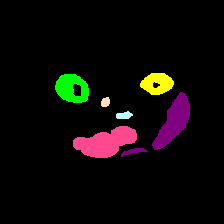}
    \end{minipage}%
    \begin{minipage}{0.23\hsize}
        \centering
        \includegraphics[scale=0.34, angle=180]{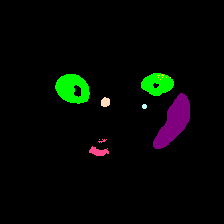}
    \end{minipage}%
    \begin{minipage}{0.23\hsize}
        \centering
        \includegraphics[scale=0.34, angle=180]{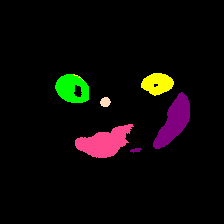}
    \end{minipage}%
    \end{tabular}
        \begin{tabular}{cccc}
      \begin{minipage}{0.23\hsize}
        \centering
        \includegraphics[scale=0.34, angle=180]{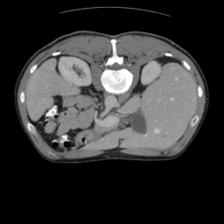}
        \subcaption{Input image}
    \end{minipage}%
    \begin{minipage}{0.23\hsize}
        \centering
        \includegraphics[scale=0.34, angle=180]{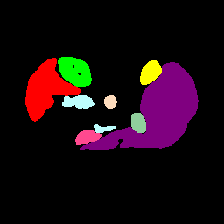}
        \subcaption{Ground truth}
    \end{minipage}%
    \begin{minipage}{0.23\hsize}
        \centering
        \includegraphics[scale=0.34, angle=180]{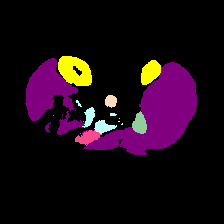}
        \subcaption{Dice loss}
    \end{minipage}%
    \begin{minipage}{0.23\hsize}
        \centering
        \includegraphics[scale=0.34, angle=180]{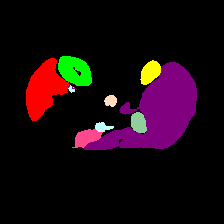}
        \subcaption{Our loss}
    \end{minipage}%
    \end{tabular}
    \caption{Qualitative comparison based on visualization. (a) Input image, (b) ground truth, (c) Dice loss, and (d) Adaptive t-vMF Dice loss ($\lambda = 256$).}
\end{figure*}
\subsection{Datasets}
We used the CVC-ClinicDB \cite{b17} and Kvasir-SEG \cite{b18} datasets for binary segmentation.
The CVC-ClinicDB dataset contains 612 RGB colonoscopy images with labeled polyps from MICCAI 2015 having a pixel resolution of $288 \times 384$. 
The Kvasir-SEG dataset contains 1,000 polyp images with a pixel resolution ranging from $332 \times 487$ to $1,920 \times 1,072$ and their corresponding ground truth.
Following \cite{b19}, we used an 80\%, 10\%, and 10\% split for training, validation, and testing.

For multi-class segmentation, we used the Automated Cardiac Diagnosis Challenge (ACDC) \cite{b22} and Synapse multi-organ segmentation \cite{b23} datasets.
The ACDC collects examination results from different patients, acquired from MRI scanners, found in the MICCAI 2017 dataset.
Each patient scan is manually annotated for the left ventricle (LV), right ventricle (RV), and myocardium (MYO).
Following \cite{b9}, we split the data into 70 training cases, 10 validation cases, and 20 test cases.
The Synapse multi-organ segmentation dataset is from the Multi-Atlas Abdomen Labeling Challenge with a total of 3779 axial contrast-enhanced abdominal clinical CT images from the MICCAI 2015 dataset.
We used 30 abdominal CT volumes, where each CT volume consists of 85 to 198 slices having a pixel resolution of $512 \times 512$.
Following \cite{b9}, we evaluated 8 abdominal organs (aorta, gallbladder, spleen, left kidney, right kidney, liver, pancreas, spleen, and stomach) with a split of 18 training, 6 validation, and 6 test cases.

\subsection{Training conditions and evaluation metrics}
For comparison, we used Dice loss \cite{b10}, Generalized Dice loss \cite{b11}, Noise-robust Dice loss \cite{b12}, WS Dice loss \cite{b13}, Focal Dice loss \cite{b14}, $\beta FL-log (Dice)$ loss \cite{b15}, and Focal Tversky loss \cite{b16}.
In addition, U-Net \cite{b21} with full-scratch training and TransUNet \cite{b9} were applied as the networks.
The batch size was set to 24, the number of training iterations was set to 4,000, and the optimizer was a stochastic gradient descent (SGD) with a momentum of 0.9 and weight decay of $2\times10^{-4}$.
We aligned a learning schedule using the rule described in \cite{b9}.
The initial learning rate was 0.01, and we decayed by $lr \times (1 - iteration/iteration_{max})^{0.9}$, where $lr$ is a learning rate and $iteration_{max}$ is the maximum number of iterations.
For the data pre-processing, training samples were resized to $224 \times 224$, flipped horizontally, rotated with an angle randomly selected within $\theta$ = -90° to 90°, and normalized from zero to one.
For the inference, the images were resized to $224 \times 224$ and normalized from zero to one.

All experiments were conducted using a five-fold cross validation and we applied the DSC as an evaluation metric.
We utilize a single Nvidia RTX Quadro 8000 GPU as a calculator.

\subsection{Results}
Table 1 shows the average DSC on test images in the CVC-ClinicDB and Kvasir-SEG datasets.
The red letters show the best DSC.
For the CVC-ClinicDB dataset, t-vMF Dice loss with $\kappa = 128$ was improved by over 6.14\% in comparison with the original Dice loss when we used U-Net, and Adaptive t-vMF Dice loss with $\lambda = 128$ was improved by over 6.66\%.
In the Kvasir dataset, the average DSC was improved by over 2.11\% when we used t-vMF Dice loss with $\kappa = 128$ for U-Net in comparison with the original Dice loss.
Although the proposed losses are a simple expansion of the Dice loss, they are effective in comparison with conventional loss functions for a semantic segmentation.

Table 2 shows the evaluation results on test images in the ACDC dataset.
Although conventional loss functions have a smaller DSC than Dice loss, our proposed loss function has the highest average DSC for U-Net and TransUNet.
In the case of t-vMF Dice loss, a large $\kappa$ gave a lower average DSC.
However, Adaptive t-vMF Dice loss further improves the DSC, even if the parameter $\lambda$ is larger and the loss function has a higher compact similarity.
These results demonstrate that there is an appropriate $\kappa$ for each class.

Table 3 shows the evaluation results on the test images, and Fig. 2 presents a qualitative comparison of different approaches through a visualization of the Synapse multi-organ segmentation dataset.
Adaptive t-vMF Dice loss achieved the highest average DSC with $\lambda = 256$ for U-Net and $\lambda = 32$ for TransUNet.
Compared to the original Dice loss, we confirmed an improvement of over 3.65\% for U-Net and over 1.81\% for TransUNet.
In Fig. 2, the incorrect prediction of the green class (left kidney) as the yellow class (right kidney) and the red class (Spleen) as the purple class (Liver) was improved, and a correct recognition was achieved.
We confirmed that our loss function predicted fewer false positives and maintained finer information.

\section{Conclusion}

In this paper, we showed that the original Dice loss can be rewritten, and introduced a novel t-vMF Dice loss using the t-vMF similarity.
Furthermore, we presented Adaptive t-vMF Dice loss that automatically determines the parameter $\kappa$ based on the validation accuracy, and it was possible to use more compact similarities for easy classes and wider similarities for difficult classes.
As demonstrated through experiments conducted on four datasets, t-vMF Dice loss and Adaptive t-vMF Dice loss showed significantly improved accuracy in comparison with conventional loss functions.
However, the accuracy of the class for small regions remains insufficient, and therefore improving the accuracy of these classes is one of our future study.

%
%
%
%

\end{document}